\documentclass[iop]{emulateapj}
\usepackage{epsfig}

\usepackage[usenames,dvips]{color}

\shorttitle{Discovery of an ultracompact millisecond pulsar binary candidate}
\shortauthors{Kong et al.}

\newcommand{\chandra}{{\it Chandra}}

\newcommand{\fermi}{{\it Fermi}}

\newcommand{\msp}{2FGL\,J1653.6−-0159}

\begin{document}
\slugcomment{Accepted for publication in ApJL}
\title{Discovery of an ultracompact gamma-ray millisecond pulsar binary candidate}

\author{Albert~K.~H. Kong\altaffilmark{1}, Ruolan Jin\altaffilmark{1}, T.-C. Yen\altaffilmark{1}, C.-P. Hu\altaffilmark{2}, C.~Y. Hui\altaffilmark{3}, P.~H.~T. Tam\altaffilmark{1}, J. Takata\altaffilmark{4},  L.~C.~C. Lin\altaffilmark{1}, K.~S. Cheng\altaffilmark{4}, S.~M. Park\altaffilmark{3}, C.~L. Kim\altaffilmark{5}
}

\affil{$^1$ Institute of Astronomy and Department of Physics, National Tsing Hua University, Hsinchu 30013, Taiwan; akong@phys.nthu.edu.tw}
\affil{$^2$ Graduate Institute of Astronomy, National Central University, Jhongli 32001, Taiwan }
\affil{$^3$ Department of Astronomy and Space Science, Chungnam National University, Daejeon, Republic of Korea} 
\affil{$^4$ Department of Physics, University of Hong Kong,
Pokfulam Road, Hong Kong} 
\affil{$^5$ Department of Physics and Astronomy, Seoul National University, Republic of Korea}



\begin{abstract}
We report multi-wavelength observations of the unidentified \fermi\ object \msp. With the help of high-resolution X-ray observation, we have identified an X-ray and optical counterpart of \msp. The source exhibits a periodic modulation of 75 min in optical and possibly also in X-ray. We suggest that \msp\ is a compact binary system with an orbital period of 75 min. Combining the gamma-ray and X-ray properties, \msp\ is potentially a black widow/redback type gamma-ray millisecond pulsar (MSP). The optical and X-ray lightcurve profiles show that the companion is mildly heated by the high-energy emission and the X-rays are from intrabinary shock. Although no radio pulsation has been detected yet, we estimated that the spin period of the MSP is $\sim 2$ ms based on a theoretical model. If pulsation can be confirmed in the future, \msp\ will become the first ultracompact rotation-powered MSP.
\end{abstract}


\section{Introduction}
The {\it Fermi Gamma-ray Space Telescope (Fermi)} has revolutionized our understanding of the high-energy universe. In particular, gamma-ray emitting pulsar is a major population discovered with the Large Area Telescope (LAT) onboard \fermi. 
In the second LAT pulsar catalog, there are 117 gamma-ray pulsars for which 43 of them are millisecond pulsars (MSPs; Abdo et al. 2013; Ray et al. 2012).
MSPs are of particular interest because they represent an important stage for the evolution of compact stars.  In recent radio surveys of \fermi-observed MSPs,
$> 75$\% them are found in binary systems (see Abdo et al. 2013 for references) and some of the binary systems have a tight orbit ($< 24$ hours). Based on the radio and
optical lightcurves, a significant amount of intrabinary materials exist in the systems and it is very likely that the gamma-ray radiation from the pulsar and/or the pulsar wind is ablating the companion. Eventually it will
evaporate the companion leaving an isolated MSP. Since the pulsar is destroying its companion, it is called the black widow (if the companion star mass $M_c < 0.05M_\odot$) or redback ($M_c > 0.1M_\odot$) pulsar. 
With new radio pulsar surveys targeted on \fermi\ sources,
new black widows and redbacks have been found (e.g. Roberts 2013; Ray et al. 2012; Abdo et al. 2013). 

Traditionally, MSPs are discovered and studied with radio timing. Indeed, all 43 gamma-ray emitting MSPs are ``radio-loud'' (Abdo et al. 2013). However, ``radio-dim'' MSPs have not been identified so far. This may be because gamma-ray emissions 
from MSPs always accompany radio emissions; the radio beam must be large enough so that an observer can see the radio emission in any geometrical configurations.
On the other hand, radio and gamma-ray emission regions can be different and depending on the geometry, it can result in ``radio-dim'' MSPs if we miss the radio beam (e.g., Venter et al. 2009). If MSPs are ``radio-dim'', radio observations may not be able to find them. Alternatively, gamma-ray observations are the best way to identify this class of sources. Interestingly, nearly 1/3 of the 1873 \fermi\ gamma-ray sources are still unidentified (Nolan et al. 2012) and they are the best candidates for ``radio-dim'' MSPs. 

To identify suitable targets for investigation, we first selected candidates from the  \fermi\ unidentified source catalog based on three criteria: 1) source variability; 2) high Galactic latitude, and 3) gamma-ray spectral shape. We used the variability index in the \fermi\ catalog to characterize source variability. For pulsars, we expect that they are steady sources. We also identified potential candidates from the gamma-ray spectra.  For gamma-ray pulsars, their gamma-ray spectra are usually described by a power-law plus an exponential cutoff model (Abdo et al. 2010). We selected sources that are not well fitted with a power-law model as shown in the catalog. We have carried out a multi-wavelength campaign to search for such objects (Kong et al. 2012; Hui et al. 2014a) and identified the first ``radio-dim'' MSP candidate 2FGL\,J2339.6--0532 (Kong et al. 2011; Romani \& Shaw 2011; Kong et al. 2012). However, more recently, radio and gamma-ray pulsation (Ray et al. 2014) as well as radio continuum emission (Kong et al. 2013) were discovered and the source is no longer a ``radio-dim'' MSP.

In this Letter, we report a multi-wavelength identification of a ``radio-dim'' gamma-ray MSP candidate that could be associated with an ultracompact X-ray binary.

\section{Multi-wavelength Identification}
\msp\ is one of the bright \fermi\ LAT sources found in the first three months of \fermi\ operation (Abdo et al. 2009) and it remains unidentified. In the second \fermi\ LAT source catalog (2FGL; Nolan et al. 2012), \msp\ has a curvature significance of 5.3 indicating that the spectral shape is significantly curved and a variability index of 17 which is equivalent to a steady source (see Section 3.3 for a detailed analysis using 5.8 years of LAT data). Finally, \msp\ is located at a Galactic latitude of $25^\circ$. All these properties indicate that \msp\ is a gamma-ray MSP candidate. Because of these, deep radio timing observations for pulsation searches have been carried out
extensively. However, no pulsation has been detected yet (Ray et al. 2012; Barr et al. 2013).

Using the same technique as for searching the X-ray/optical counterpart of the first ``radio-dim" MSP candidate 2FGL\,J2339.6--0532, we first checked the archival X-ray data to look for possible X-ray counterparts in the \fermi\ error circle of \msp. The field of \msp\ was observed with \chandra\ on 2010 January 24 for 21 ks with ACIS-I. We reprocessed the data with updated calibration files. Within the 95\% \fermi\ error circle, there is only one relatively bright X-ray source (CXOU J165338.0-015836). The X-ray-to-gamma-ray flux ratio is about 0.007 while all other \chandra\ sources in the error circle are much fainter (see Cheung et al. 2012) with their X-ray-to-gamma-ray flux ratios less than 0.02\%. Such a low flux ratio is not typical for an X-ray counterpart to a \fermi\ source.

We tentatively identified CXOU J165338.0--015836 as the possible X-ray counterpart to \msp. This X-ray source was also listed in Cheung et al. (2012) as a potential X-ray counterpart to \msp. Within the 90\% error circle of CXOU J165338.0--015836, we identified a $R\sim20$ star that is 0.44 arcsec from the \chandra\ position in the SuperCOSMO Sky Survey (Hambly et al. 2001). This is also the same star identified in the USNO catalog (Cheung et al. 2012). Given the positional coincidence, we suspected that this is the optical counterpart to CXOU J165338.0-015836. Based on the X-ray spectral fit from Cheung et al. (2012), CXOU J165338.0-015836 has an unabsorbed 0.5--2 keV flux of $9.5\times10^{-14}$ erg s$^{-1}$ cm$^{-2}$. This yields an X-ray-to-optical flux ratio of about 3 which is too high for a foreground star and is not typical for an AGN (e.g. Green et al. 2004; Laird et al. 2009). Instead, the $B-R$ color ($B-R \approx 0.5-1$ based on the USNO catalog) looks like a late-type star. 
All these are very similar to 2FGL\,J2339.6--0532 (Romani \& Shaw 2011; Kong et al. 2012). We therefore believe that this X-ray/optical counterpart is associated with \msp\ as a gamma-ray MSP candidate.

We have searched for the NVSS
catalog and there is no 1.4 GHz source at the X-ray/radio position. Deep radio timing observations have been conducted to look for pulsation from \msp\ and a flux limit of 70$\mu$Jy (1.36 GHz) has been set (Barr et al. 2013). 

\section{Observations and Data Analysis}

Because \msp\ is a potential ``radio-dim'' gamma-ray MSP, we have performed an optical follow-up observation for its optical counterpart. We also reanalyzed the \chandra\ X-ray data and performed a detailed analysis using 5.8 years of \fermi\ LAT data.

\subsection{Optical Photometry}

We carried out an optical time-series observation for the proposed optical counterpart of \msp\ in $r'$ and $g'$-bands with the 2.5m Isaac Newton Telescope (INT) and the Wide Field camera (WFC) in La Palma on June 4-5, 2014. The exposure time for all observations is 6 minutes lasting for about 6 hours each night.  The WFC was operated in a $5'\times5'$ windowing mode with a readout time of about 6 sec. All images are flat-field and bias corrected and we performed relative photometry by comparing with several comparison stars in the field. The observing time was barycentric corrected.
The lightcurves clearly show variability on a timescale of 1.2 hours (see Fig. 1). We then performed a timing analysis by using the Lomb-Scargle periodogram and a 75-min periodicity is highly significant in both bands. We then applied an observed-minus-calculated (O--C) diagram analysis to obtain a best-fit ephemeris of BJD $(2456813.48210\pm0.00078) + (0.051930 \pm 3.1 \times 10^{-5}) \times N$. The phase 0 is defined as the optical maximum in the $r'$-band (superior conjunction; see Discussion). We also show a binned $g'-r'$ lightcurve that also exhibits a 75-min periodicity in Fig. 1.

\begin{figure}
\centerline{
\epsfig{figure=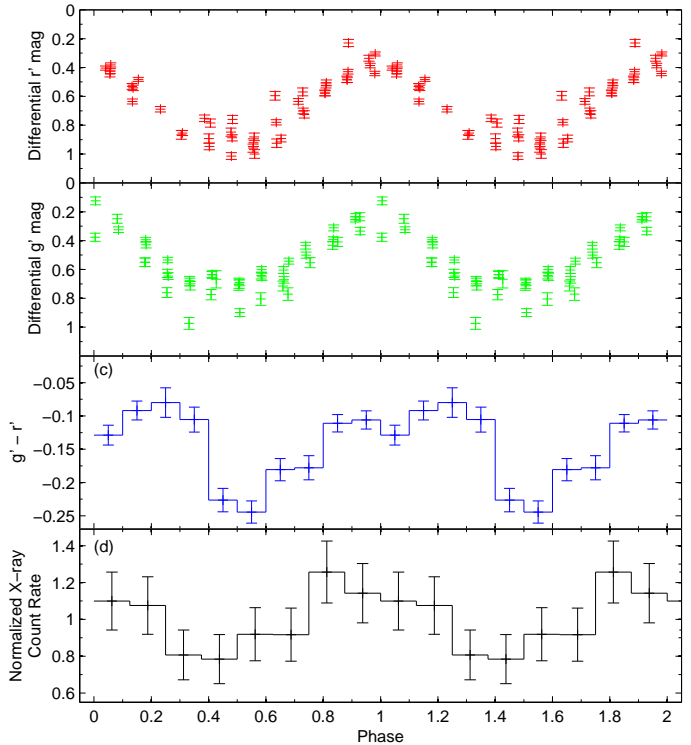,width=10cm}}
      \caption{Folded lightcurve of $r'$-band (panel a), $g'$-band (panel b), binned $g'-r'$ (panel c), and  \chandra\ (panel d) observations of \msp\ with a best-fit period of 74.7792 min. The phase zero is defined as the $r'$-band optical maximum (BJD 2456813.48210).  It is evident that both optical and X-ray lightcurves show similar modulation.}
      
\end{figure}

\subsection{\chandra}
The phase-averaged \chandra\ observation was reported in Cheung et al. (2012) and here we focus on the timing analysis and phase-resolved spectroscopy. A detailed X-ray analysis will be present in Hui et al. (2014a). Using the optical timing ephemeris above, we applied a barycentric correction to the X-ray photon arrival time and folded the background subtracted 0.3--8 keV lightcurve. A 75-min periodicity is suggestive (Fig. 1) and is significant at 97\% level based on an $H$-statistic ($H$-value=8.65). A more sensitive X-ray observation is required to confirm this.

To investigate whether the X-ray spectral properties vary at different orbital phases, we have performed 
a phase-resolved spectroscopy.  We have extracted the spectra from two phase ranges that encompass the 
peak ($\phi=0.0-0.4$) of the X-ray orbital modulation and the other coveres the trough ($\phi=0.5-0.8$). 
The background spectra for the corresponding temporal coverage were sampled from a nearby source-free 
region. The extraction of the source and background as well as the generation of response files were 
done by using the CIAO script {\it specextract}. After background subtraction, there were $\sim155$~cts 
and $\sim72$~cts from the peak and trough intervals in $0.3-8$~keV respectively. We grouped the 
spectra so as to have at least 10 counts per spectral bin. 

Since Hui et al. (2014a) has confirmed the X-rays from 2FGL~J1653.6-0159 are non-thermal dominant, 
we examined both spectra with an absorbed power-law model (cf. Hui et al. 2014b). In view of small photon 
statistics, we fixed the column absorption at the value inferred from the phase-averaged analysis, 
i.e. $N_{H}=10^{21}$~cm$^{-2}$ (cf. Hui et al. 2014a). For the peak interval, the best-fit yields a photon index of 
$\Gamma=1.6\pm0.2$ and a normalization of $1.8\pm0.3\times10^{-5}$~photons~keV$^{-1}$~cm$^{-2}$~s$^{-1}$ at 1 keV.
For the trough interval, the corresponding parameters are $\Gamma=1.6\pm0.3$ and 
$8.7^{+2.1}_{-1.9}\times10^{-6}$~photons~keV$^{-1}$~cm$^{-2}$~s$^{-1}$ at 1 keV. 
Based on this observation, the photon indices inferred in these two intervals are consistent. 
We concluded that there is no evidence of X-ray spectral variability across the orbit of 
2FGL~J1653.6-0159 found in our investigation. However, in this phase-resolved analysis, the limited photon statistics do not allow us to examine whether the spectral feature at $\sim3.5$~keV identified by Hui et al. (2014a) in a phase-averaged analysis exists or not. 

\subsection{\fermi\ LAT}

The \fermi\ LAT data used in this work were obtained between 2008 August 4 and 2014 May 30, available at the Fermi Science Support Center\footnote{\url{http://fermi.gsfc.nasa.gov/ssc/}}. We used the Fermi Science Tools v9r33p0 package to reduce and analyze the data. Only reprocessed pass 7 data classified as ``source'' events arriving at zenith angles $<$100$^\circ$ were used. The instrument response functions ``P7REP\_SOURCE\_V15'' were used.

We carried out a binned maximum-likelihood analysis using \emph{gtlike} of 0.1--300~GeV events from the rectangular region of 21$^\circ\times$21$^\circ$ centered at 2FGL~J1653.6$-$0159. We subtracted the background gamma-ray flux by including the Galactic diffuse model (gll\_iem\_v05\_rev1) and the isotropic background (iso\_source\_v05), as well as all sources in the second Fermi/LAT catalog (Nolan et al., 2012) within the circular region of 25$^\circ$ radius around 2FGL~J1653.6$-$0159. The recommended spectral model for each source employed in the catalog was used, while we modeled
2FGL~J1653.6$-$0159 with a simple power law (PL)
\begin{equation}
\frac{dN}{dE} = N_0 \left(\frac{E}{E_0}\right)^{-\Gamma},
\end{equation}
and a power law with exponential cutoff (PLE)
\begin{equation}
\frac{dN}{dE} = N_0 \left(\frac{E}{E_0}\right)^{-\Gamma}\exp(-\frac{E}{E_\mathrm{c}}).
\end{equation}
The normalization values were set free for the Galactic and isotropic diffuse background, as well as sources within 10$^\circ$ from 2FGL~J1653.6$-$0159.

\begin{figure}
\centerline{
\epsfig{figure=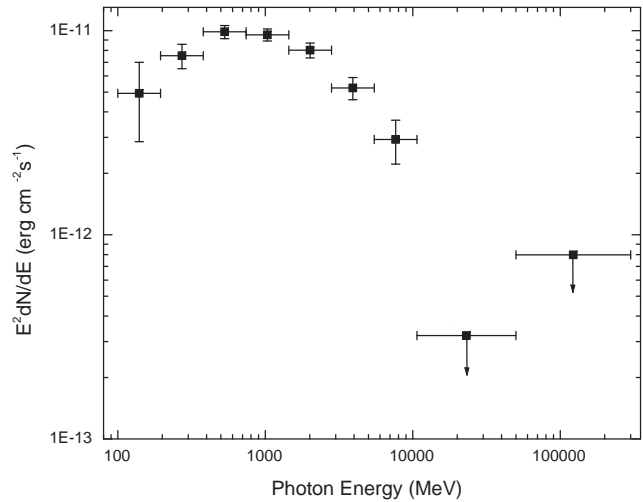,width=10cm}}
      \caption{The gamma-ray spectrum derived using 5.8-year \fermi\ LAT data (see Section 3.3 for details).}
      \label{gamma_spec}
\end{figure}

Based on the difference between $-$log(likelihood) values for both models, the PLE model is preferred over the PL model by 10$\sigma$. Using the PLE model, $\Gamma=-1.7\pm0.1$, $E_\mathrm{c}=3.3\pm0.5$~GeV, and a 100~MeV--300~GeV energy flux of $(3.52\pm0.14)\times10^{-11}$~erg~cm$^{-2}$~s$^{-1}$ are obtained.
These spectral parameter values are typical of those found for gamma-ray millisecond pulsars (Abdo et al. 2009). We divided the 100~MeV--300~GeV gamma-rays into 9 energy bins and reconstructed the flux using \emph{gtlike} for each band separately. The spectrum is shown in Fig.~\ref{gamma_spec}. For each bin above 10~GeV, a PL model with $\Gamma_\gamma=$3.0 was assumed for 2FGL~J1653.6$-$0159 to derive the 90\% confidence-level upper limits.

We probed any long-term flux variation by constructing a gamma-ray light curve from August 2008 to May 2014 with a bin size of 90 days. The source model as described above was used to estimate the background using likelihood analysis. No significant variation was seen in the light curve. Folding the gamma-ray photons at the 75-min orbital period did not reveal any gamma-ray modulation as well. We derived a $3\sigma$ upper limit of 21\% for the 1--10 GeV fractional variation of the 75-min modulation (see de Jager 1994).

\section{Discussion}

Using optical, X-ray, and gamma-ray data, we identified the X-ray and optical counterpart of the \fermi\ unidentified source, \msp. The most interesting feature is the 74.7792-min periodicity found in optical data and possibly in X-ray data (see Fig. 1). This period can be considered as the binary period of a compact binary system. Given the gamma-ray spectrum  and the X-ray-to-gamma-ray flux ratio are consistent with a typical gamma-ray pulsar, we propose that \msp\ is a black widow/redback MSP. Since deep radio timing has not yet found the pulsation (Ray et al. 2012; Barr et al. 2013), \msp\ is very likely a ``radio-dim'' gamma-ray MSP. This system is very similar to 2FGL\,J2339.6--0532. While it is no longer a ``radio-dim'' MSP, it proves that this identification technique can discover MSP candidates from unidentified gamma-ray sources. Using this technique, PSR J1311--3430 was discovered (Romani 2012) and more recently 1FGL\,J0523.5--2529 was identified as a probable gamma-ray pulsar without a radio counterpart (Strader et al. 2014).

Like 2FGL\,J2339.6--0532 and other black widows and redbacks, the optical emission of \msp\ is affected by the high-energy heating from the MSP on the companion surface, producing orbital modulation. This indicates that the optical maximum corresponds to the superior conjunction at which the MSP is in between the companion and the observer. We also did a cross-correlation analysis of all lightcurves and did not find any significant time lag. It is worth noting that the X-ray observation was taken more than 4 years ago and given our current uncertainty of the orbital period, a direct comparison may not be correct. Contemporary X-ray/optical observations are required to study the phase alignment of \msp.

 With an observed gamma-ray flux of $F_{\gamma}\sim 3\times 
10^{-11}{\rm erg~cm^{-2}~s^{-1}}$ and an assumed typical distance for gamma-ray MSP of $d\sim 1$kpc, the gamma-ray 
power of \msp\ is $L_{\gamma}=4\pi F_{\gamma}d^2
\sim 4\times 10^{33}f_{\Omega}d_{1kpc}^2\mathrm{erg s^{-1}}$, where 
$f_{\Omega}$ is the viewing factor. The estimated
gamma-ray luminosity $L_{\gamma}\sim 4\times 10^{33}{erg~s^{-1}}$ 
is a typical value of pulsed gamma-rays from well-known MSPs (c.f. Table~1), and it is likely originated from the 
pulsar magnetosphere. The orbital modulating X-ray emissions will be produced 
by the intrabinary shock due to the interaction between the 
pulsar wind and the stellar wind (Kong et al. 2012). The inferred X-ray luminosity 
from the observed flux is $L_{X}\sim 2\times 10^{31} d_{1kpc}^2 {\rm erg~s^{-1}}$, which 
is in the range of the observed X-ray luminosity of known black widow/redback
 systems (see Table~1). 

Since the efficiency of the gamma-ray luminosity of MSPs 
is in general $10\%$ (Abdo et al. 2013),  suggesting the spin down power 
is of order of $L_{sd}\sim 4\times 10^{34} f_{\Omega}d^2_{1kpc}{\rm erg~s^{-1}}$. 
Assuming a typical dipole  magnetic field of MSPs, 
$B_s\sim 10^8$G, the rotation period is estimated as 
$P\sim 1.8{\rm ms}(B_s/10^8{\rm G})^{1/2}f_{\Omega}^{-1/4}d_{1kpc}^{-1/2}$ (Takata, Cheng \& Taam 2012). 

\begin{table*}
\begin{center}
\caption{Physical parameters of \msp\ and 6 other representative black widow/redback systems.}
\begin{tabular}{ccccccl}
\hline\hline
 Source & D & $P$ & $L_{sd}$ & $L_{\gamma}$ & $L_{X}$ & $P_{ob}$ \\
 & (kpc) & (ms) & ($10^{34}$erg/s) & ($10^{33}$erg/s) & ($10^{31}$erg/s) & (day) \\
\hline
\msp\ & 1$^a$ & ? & ? & 4 & 2 & 0.052 \\
1FGL J2339.7-‐0531 & 0.7 & 2.88 & 2 & 2 & 4 & 0.19 \\
PSR J1311--3430 & 1.4 & 2.56 & 4.9 & 15 & 5.6& 0.065\\
PSR B1957+20 & 2.5 & 1.61 & 10 & 15 & 45 & 0.38 \\
PSR J1023+0038 &1.3 & 1.69 & 5 & 1.2 & 10 & 0.2 \\
PSR J1723--2837 & 0.75 & 1.86 & 5 & 1 & 7 & 0.62 \\
PSR J2051--0827 & 1 & 4.51 & 0.5 & 0.7 & 0.1 & 0.1 \\
\hline \hline
\end{tabular}
\end{center}
\par
\smallskip
\hspace{3.2cm}$^a$ Assumed distance \\
\end{table*}

We do not find significant difference between different orbital phase-resolved
 X-ray spectra. In the context of intrabinary shock between the pulsar wind and 
the stellar wind from the companion star, this might imply that the orbit is almost circular so that the shock distance 
from the pulsar has little dependence on the orbital phase. And hence the spectral properties do not modulate with the 
orbital phase.

  We expect that the optical modualtion is caused by the irradiation 
of the pulsar emissions, indicating the optical peak phase corresponds to the 
superior conjunction. We also expect that the orbital modulation of 
the X-ray emission is caused by 
the Doppler boosting effects of the post-shocked pulsar wind. Since 
the shift between the optical peak and X-ray peak is less than 0.5 orbital phase, 
we expect that the shock covers the pulsar, which will be similar to 
the shock geometry of the redback PSR\,J1023+0038 (Takata et al. 2014; Li et al. 2014). 
The non-detection of the gamma-ray orbital period may be because the amplitude of the orbital modulation is too small to be detected by \fermi\ and/or the inverse-Compoton flux of pulsar wind is smaller than the magnetospheric emissions.

Black widows/redbacks are likely in the late stages of recycling, providing a crucial link between MSPs and accreting millisecond X-ray pulsars (AMXPs). In the family of AMXPs, there are a few ultracompact systems with orbital periods less than 80 min (XTE\,J1751--305, XTE\,J0929--314, XTE\,J1807--294, HETE\,J1900.1--2455, Swift\,J1756.9-2508, NGC 6440 X--2) and therefore ultracompact MSPs should be expected. Indeed, simulations suggest that there should be a large number of ultracompact MSPs in globular clusters (e.g. Rasio et al. 2000). Although the formation of MSPs in the field might be different from those in clusters (e.g., Belczynski \& Taam 2003), some may be formed in clusters and later kicked out to distribute in the field.
The striking feature of \msp\ is its very compact binary orbital period (74.7792 min) and it may be the first of its kind. If future radio or gamma-ray observations confirm its MSP nature, it will be the most compact rotation-powered MSP binary ever found. Interestingly, the short orbital period can be linked with ultracompact X-ray binaries for which the binary system consists of a compact object (neutron star or black hole) and a degenerate or partially degenerate companion star under a compact ($< 80$ min; defined by the period distribution of X-ray binaries) binary orbit (e.g. Nelson et al. 1986; Podsiadlowski et al. 2002). These systems are usually hydrogen-poor objects and if \msp\ is associated with an ultracompact system, the companion is likely a helium star or a white dwarf. Based on binary evolution calculations, \msp\ is more likely a black widow system with a companion much less than 0.1 $M_\odot$ (Chen et al. 2013). Recently, a new black widow system, PSR\,J1311−-3430, was found to have an orbital period of 94 min and the companion is a helium rich star (Romani et al. 2012). It may be suggestive that the companion star of \msp\ could also be helium rich similar to ultracompact X-ray binaries. Future optical spectroscopy will confirm this.

Last but not least, pulsation search is required to confirm the MSP nature and gamma-ray pulsation search with LAT will be the best way to try given that we now have the binary orbital period. At the same time, deep radio imaging will prove whether the source is ``radio-dim'' or not because imaging observations will not be affected by scattering/absorption due to the intrabinary environment.

Note to readers: Another paper by Romani et al. (2014) was submitted slightly earlier than ours and they also identify a similar orbital period. With optical spectroscopy, they found that the companion star is hydrogen poor.

The INT is operated on the island of La Palma by the Isaac Newton Group in the Spanish Observatorio del Roque de los Muchachos of the Instituto de Astrofísica de Canarias. We thank Diego Torres and Nanda Rea to arrange the INT observations, and Ovidiu Vaduvescu and Teo Mocnik for helping the observing runs.
AKHK, RJ, and TCY are supported by the Ministry of Science and Technology
 of the Republic of China (Taiwan) through grants 100-2628-M-007-002-MY3, 100-2923-M-007-001-MY3, and 103-2628-M-007-003-MY3.
JT and KSC are supported by a 2014 GRF grant of Hong Kong Government under HKU 17300814P.  PHT and CPH are supported by the Ministry of Science and Technology of the Republic of China (Taiwan) through grants 101-2112-M-007-022-MY3 and 102-2112-M-008-020-MY3, respectively. CYH is supported by the National Research Foundation of Korea through grant 2011-0023383. 

{\it Facilities:} \facility{Isaac Newton Telescope}, \facility{Chandra}, \facility{Fermi}

\end{document}